\documentclass[preprints,article,accept,pdftex,moreauthors]{Definitions/mdpi} 
\usepackage{ulem}
\usepackage{comment}
\firstpage{1} 
\pubvolume{1}
\issuenum{1}
\articlenumber{0}
\pubyear{2026}
\copyrightyear{2026}
\datereceived{ } 
\daterevised{ } 
\dateaccepted{ } 
\datepublished{ } 

\Title{Geometric Cosmology Models: Statistical Analysis with Observational Data}

\Author{Matías Leizerovich$^{1,2}$\orcidA{}, Luisa G. Jaime$^{3,4}$\orcidB{}, Susana J. Landau $^{2}$\orcidC{},  and Gustavo Arciniega$^{3}$\orcidD{}}

\AuthorNames{Matías Leizerovich, Susana J. Landau, Luisa G. Jaime and Gustavo Arciniega}

\address{
$^{1}$ \quad Universidad de Buenos Aires, Facultad de Ciencias Exactas y Naturales, Departamento de Física. Buenos Aires, Argentina. \\
$^{2}$ \quad CONICET - Universidad de Buenos Aires, Instituto de Física de Buenos Aires (IFIBA). Buenos Aires, Argentina \\
$^{3}$ \quad Departamento de F\'{\i}sica, Facultad de Ciencias, Universidad Nacional Autónoma de México, 04510 CDMX, México\\
$^{4}$ \quad Instituto Galego de F\'{\i}sica de Altas Enerxías (IGFAE), Universidade de Santiago de Compostela, E-15782 Santiago de Compostela, Spain
}

\corres{Correspondence: mleize@df.uba.ar}

\abstract{Although the standard cosmological model successfully describes most current observational data, it faces several theoretical and observational challenges that motivate the exploration of alternative frameworks. In this work, we investigate a class of geometric cosmology models  (GC) obtained by adding an infinite tower of higher-order curvature invariants to the Einstein--Hilbert action. Focusing on an exponential ansatz for  the characteristic function entering the modified Friedmann equations, we derive
 the late-time background evolution for three families of solutions 
within this framework, named as : i) GILA, ii) GR-deformation, iii)Non-GR contribution. These models are confronted with recent Cosmic 
Chronometer and Type Ia supernova data, as well as age estimates of the 
oldest globular clusters - a constraint frequently overlooked in the 
literature. The stiffness of the equations in certain regions of 
parameter space, together with technical difficulties arising from the 
inclusion of the globular cluster bound, motivates the development of a 
dedicated methodology as an alternative to standard Markov Chain Monte 
Carlo techniques. Our results show that two entire families of GC models (Non-GR contribution and GR-deformation) are ruled out by the data,  whereas some families within the GILA model can successfully account for all data sets. For these models,  meaningful constraints on their free parameters can be derived from the statistical analysis. Nevertheless, model comparison criteria reveal a preference in the data for $\Lambda$CDM over the GILA models examined here.
Although none of the proposed models provides a preferred alternative to $\Lambda$CDM given the specific characteristic function considered here, this work establishes a clear methodology for testing alternative cosmological models, including the globular cluster constraint and indicates the way for future research of GILA models with alternative choices of the characteristic function.}

\begin{document}
\section{Introduction} \label{sec:outline}

The discovery of the late-time acceleration of the Universe from observations of type Ia supernovae (SNIa), posed an important challenge to theoretical cosmology: to establish the physical mechanism responsible for this acceleration. The proposal of the standard cosmological model (called $\Lambda$CDM, which assumes General Relativity as the theory that describes the gravitational interaction) to solve this issue is to add a cosmological constant to Einstein's equations. However, this proposal faces some theoretical and observational issues that have been discussed in the literature. Among the theoretical issues, the most relevant one is that  the value of the observed cosmological constant has a difference with the theoretical prediction of about 60 or 120 orders of magnitude, if one considers that the cosmological constant arises from the vacuum energy of standard model fields. On the other hand, there are tensions between the values of cosmological parameters inferred with different data sets. The most prominent example is the so-called "Hubble tension" which refers to a more than $5\sigma$ tension between the value of the Hubble constant $H_0$ inferred from Cosmic Microwave Background data from the Planck collaboration \citep{Planckcosmo2018} and the one obtained with Supernovae and Cepheid data \citep{Riess2022}. Also, recent results from the DESI collaboration \citep{DESI:2025zgx}, suggest that the dark energy equation of state has a different behavior from the one predicted by the $\Lambda$CDM  model.  Given that the origin of the observed late-time acceleration remains an open question, and the persistent tensions among inferred cosmological parameters, there is strong motivation to explore alternative cosmological models. The proposals discussed in the literature can be grouped into two families. i) dark energy, where the late time acceleration of the universe is driven by a either a fluid \citep{darkenergy1,darkenergy2} or a scalar field \citep{quintessence_exp_pot2} ii) modified gravity, where the theory that describes gravitation is no longer General Relativity, but an alternative theory of gravity \citep{review_mod_grav, alt_grav_1}.


 In this paper, we analyze a theory of gravity that introduces an infinite tower of higher-order curvature scalars into the modified Einstein–Hilbert action. The original proposal \citep{Arciniega:towards} was formulated by adding cubic-order terms to the gravitational action. It was shown that the inclusion of these terms satisfies the so-called Einsteinian conditions \citep{Bueno:2016xff,Bueno:2016ypa}, yielding second-order equations in the case of a homogeneous and isotropic universe and, moreover, giving rise to an early stage of accelerated expansion. Subsequently, in  \citet{Cisterna:2018tgx}, it was shown that similar results hold upon incorporating fourth- and fifth-order curvature scalars. Finally, in Refs \citep{Arciniega:GI,Arciniega:ensayo}, this result was extended to an infinite tower of curvature scalars which, upon imposing both the Einsteinian and cosmological conditions, can produce an inflationary stage in an exponential manner, including a graceful exit to the regime of General Relativity without the need to introduce any inflaton field. In Ref. \citep{Arciniega:gabriella}, it was shown that, when the modified field equations converge to exponential functions, they yield an inflationary scenario consistent with expectations from observations of the Planck satellite \citep{Planck:2018jri}. Furthermore, in  Ref. \citep{Jaime:viability}, it was demonstrated that this gravitational theory provides an inflationary period capable of resolving the standard problems of inflation (horizon, flatness, and monopole problems). Given the results of these works, attention turned to whether this framework could also account for a viable late-time accelerated expansion consistent with observations. In Ref. \citep{Jaime:AUnified}, an exponential series similar to that used for the inflationary period was explored at low energy scales, showing that there exist cases worthy of statistical analysis. The present work investigates this viability in the case where the energy scale is relevant to the current value of the Hubble expansion rate. Therefore, the present analysis is restricted to the case in which the characteristic function entering the Friedmann equation takes an exponential form.

 Within this framework it is necessary to explore the different convergences of the infinite series at late time. According to this, we identify three cases of interest for our study, that we will name as follows: i) GILA, ii) GR-deformation, iii)Non-GR contribution \citep{Arciniega:2025viq}.
 
One of the important achievements of this paper lies in obtaining the late-time background evolution of the cosmological models derived from this novel proposal and testing them against recent data provided by SNIa and cosmic chronometers. We further incorporate constraints from the observed ages of the oldest globular clusters - a source of information that has not received sufficient attention in the literature. The stiffness of the Friedmann equation in certain regions of the parameter space, combined with technical difficulties arising from the inclusion of the age of the oldest globular clusters, prevented the use of a standard MCMC-based statistical analysis. This motivated the development of an alternative methodology based on Monte Carlo sampling. For the GILA models that can provide reasonable predictions for all observational data sets we obtain for the first time constraints on the parameter space.
 
The outline of the paper is as follows. In Section \ref{sec:theory} we briefly describe the modified gravity theories that we assume to build our cosmological models. Also we show the equations that describe the background dynamics of the universe in the context of our models. In section \ref{sec:data} we briefly describe the SNIa and CC data used in this paper. We also discuss recent data concerning the age of the globular clusters, which lead to an important constraint for the cosmological models that are tested here. The methodology of the statistical analysis performed to compare the late time evolution of the proposed cosmological model with recent data is described in Section \ref{sec:methodology}. In Section \ref{sec:Results} we show the results of the statistical analyses performed with cosmological models described in Section \ref{sec:theory} and observational data discussed in Section \ref{sec:data}. Finally, in Section \ref{sec:conclusions} we write our conclusions.
 

\section{Geometric Cosmology Theories} \label{sec:theory}

Let us consider a set of Lagrangian densities that satisfy the following properties:

\begin{itemize}
    \item[1.] The equations of motion for a maximally symmetric spacetime are of second order. In addition, no other particle propagates but the massless graviton.     
    \item[2.] The theory admits single-function solutions in a maximally symmetric space-time, including Taub-NUT/Bolt solutions.
    \item[3.] The field equations for a Friedmann-Lema\^itre-Robertson-Walker  metric (FLRW) are of second order.
\end{itemize}

If the theory satisfies 1 and 2, it is called Generalized Quasitopological Gravity (GQTG) (\citet{Bueno:2016lrh, Bueno:2016xff, Bueno:2016ypa, Bueno:2019ycr, Bueno:2022res, Moreno:2023rfl}), if it also satisfies 3, it is called Cosmological GQTG (CGQTG) (\citet{Arciniega:towards, Cisterna:2018tgx, Arciniega:ensayo, Arciniega:GI, Arciniega:towards, Arciniega:gabriella, Jaime:viability, Jaime:AUnified, Moreno:2023arp}). We will name it Geometric Cosmology (GC) in this work for short (\citet{Arciniega:2025viq}), in the sense that the accelerated epochs of the universe (inflation and late-time acceleration) are given by the geometric contribution of this proposal into the Friedmann equations. The GC theory is constructed by adding higher curvature Lagrangian densities by the contraction of curvature tensors. The gravitational action is thus an infinite tower of Lagrangian densities of higher order:

\begin{equation}\label{actionGC}
    S_{\text{GC}} = \frac{1}{2\kappa}\int d^4x  \sqrt{-g}\left[\left(1+\alpha_1\right) R + \sum^{\infty}_{i=3}\alpha_{i}\mathcal{R}^{(i)}\right] \, ,
\end{equation}

\noindent where $\kappa=8\pi G$ \footnote{In what follows we will consider $c=1$.} and $\alpha_{i}$ are couplings of the Lagrangian densities. Here $i=1$ corresponds with the Ricci curvature scalar, while $i=2$ is the Gauss-Bonnet or Lovelock-two gravity theory that is a topological term in dimension $D=4$\footnote{In 4 dimensions, this term does not contribute to the field equations and therefore we will remove any of such terms if they appear in the following development of the theory}. The $\mathcal{R}^{(i)}$, for $i\geq 3$, are Lagrangian densities constructed by contractions of $i$-Riemann tensors (\citet{Arciniega:GI, Moreno:2023arp, Arciniega:2025viq}). 

As usual in the cosmological scenario we assume a FLRW metric with null spatial curvature, $ds^2=-dt^2+a^2(t)(dr^2+r^2d\theta^2+r^2\sin^2\theta d\phi^2)$, and obtain the field equations for a perfect fluid with energy density $\rho$ and pressure $P$ (\citet{Jaime:AUnified}):

\begin{eqnarray} \label{eq: friedmann_eqs_gila}
    3F(H)&=& \kappa \rho,\\ \label{eq: friedmann_eqs_gila_2}
    -\frac{\dot{H}}{H}F'(H)&=&\kappa (\rho+P),
\end{eqnarray}

\noindent where $\dot{(\,)}=d/dt$, $H=\dot{a}/a$, $(\, )'=\partial_H (\,)$. Each Lagrangian density $\mathcal{R}^{(i)}$ contributes with a $H^{2i}$ after the variation of action (\ref{actionGC}) within the ansatz of an FLRW, resulting in an infinite power series for $F(H)$:

\begin{equation}\label{eq: fdeh}
    F(H)=(1+\alpha_1)H^2+\sum_{i=3}^\infty \alpha_{(i)}H^{2i}.
\end{equation}

The infinite power series for $F(H)$ can converge to functions of $H$ for a specific choice of $\alpha_{(i)}$. In particular, in the literature, the exponential convergent function $F(H)$ is the one that has been most explored. In the following, we will focus on the cosmological analysis of these exponential $F(H)$ models.

\subsection{GILA model $(\alpha_1 = 0)$}

Let us now introduce the GILA model, which is the acronym of Geometric Inflation and Late-time Acceleration given for the first time in \citet{Jaime:AUnified}. We will denote the characteristic function as $F_\gamma$ to distinguish it from other models. It can be shown that fixing $\alpha_1 =0$ results in (see Appendix \ref{sec:appendix} and \citet{Arciniega:2025viq}):

\begin{align} \label{eq: F_H_gila}
    F_\gamma(H) =& H^2 + \lambda L^{2(p-1)}         H^{2p}  e^{\lambda(LH)^{2q}} \nonumber \\
            &     - \beta   \tilde{L}^{2(r-1)} H^{2r}  e^{-\beta(\tilde{L}H)^{2s}} \, ,
\end{align}

\noindent  where $\lambda$ and $\beta$ are free dimensionless parameters and $L$ and $\tilde{L}$ are the energy scales (in units of $H_0^{-1}$) for the early and late time modifications to GR, respectively. In addition, as explained in Appendix \ref{sec:appendix} the exponents $p$, $q$, $r$, and $s$ must satisfy:

\begin{equation}
    p\geq 3, q\geq 1, r\geq 3, \text{ and } s\geq 1, 
\end{equation}

Finally, deriving with respect of $H$ we obtain

\begin{align}\nonumber
    F_\gamma'(H)& = 2H \Bigg\{1
                       + \lambda e^{\lambda(LH)^{2q}}        (LH)^{2(p-1)}         [p  + q \lambda (LH)^{2q}] \\
                      & - \beta   e^{-\beta(\tilde{L}H)^{2s}} (\tilde{L}H)^{2(r-1)} 
                      \\ \nonumber
                      &\times\Big[-r  + s \beta   (\tilde{L}H)^{2s}(LH)^{2q}\Big]  \Bigg\} \nonumber \,.
\end{align}

\subsection{GR deformation  $(-1<\alpha_1 < 0)$}\label{GRdeformation}

Now, we want to consider a slightly different modification of GR which is close to the GC theory. In this case, the parameter $\alpha_1\in (-1,0)$ and the action $S_{\text{GC}}$ on Eq. \ref{actionGC} can be written as

\begin{equation}\label{actionbetadef}
     S_{\beta}= \int d^4x\sqrt{-g}\left[\frac{R}{2\kappa_{\rm eff}}+ \sum^{\infty}_{i=3}\frac{\alpha_{i}}{2\kappa}\mathcal{R}^{(i)} \right].
 \end{equation}

Here $ \kappa_{\text{eff}}\equiv \kappa (1+\alpha_1)^{-1}=8\pi G_{\text{eff}}$, and $G_{\text{eff}}$ is the effective Newton constant. Therefore, $\alpha_1$ is related to the deformation of the General Relativity constant $\kappa$ and we call this model GR-deformation. The field equations derived from the action in Eq. \ref{actionbetadef} are the same as Eqs. (\ref{eq: friedmann_eqs_gila}) and (\ref{eq: friedmann_eqs_gila_2}), where, in this case:
\begin{equation}\label{F(H)}
    F_\beta(H)=\frac{\kappa}{\kappa_{\text{eff}}}H^2+\sum_{i=3}^\infty \alpha_{(i)}H^{2i}.
\end{equation}
The $F(H)$ series expansion can be arranged to converge to the following, taking $\alpha_1=\beta$ (see App. \ref{sec:appendix}):

\begin{equation}\label{eq: F_H_beta}
   F_\beta(H)= H^2+\lambda L^{2(p-1)}H^{2p}e^{\lambda(LH)^{2q}}-\beta H^2 e^{-\beta(\tilde{L}H)^{2s}},
\end{equation}

\noindent where $p$, $q$, and $s$ are constrained by the following:

\begin{equation}
    p\geq 3, q\geq 1, \text{ and } s\geq 2 \, .
    \label{desig_beta}
\end{equation}

The expression of $F'(H)$ in this case is  
\begin{align} \label{eq: F_H_prime_beta}
    F_{\beta}'(H) &= 2H \{ 1 + \lambda  e^{\lambda(LH)^{2q}} (LH)^{2(p-1)} [p  + q \lambda  (LH)^{2q}] \\[3pt]
                  &+           \beta    e^{-\beta(\tilde{L}H)^{2s}}             [-1 + s \beta (\tilde{L}H)^{2s}] \} \nonumber \,. 
\end{align}

Notice the $\beta H^2$ factor in the last term before the exponential. In the series expansion, this is the factor that will give $(1-\beta)H^2$. Finally, note that the convergent series in Eq. (\ref{eq: F_H_beta}) can be obtained from the GC series expansion taking $r=1$ in Eq. (\ref{eq: F_H_gila}), however, for the GILA model it is not allowed that $r<3$, which makes this case a new gravity model.

\subsection{Geometric Cosmology with non-GR contribution  $(\alpha_1 = -1)$ }\label{secc:noGR}

For the $\beta$-GR deformation theory, there is a special case when $\alpha_1=-1$, and $\kappa^{-1}_{\rm{eff}}\simeq0$. As a consequence, the Ricci curvature scalar disappears from the action (\ref{actionbetadef}):

\begin{equation}\label{actionpure}
     S_{\delta}= \int d^4x\sqrt{-g}\left(\sum^{\infty}_{i=3}\frac{\alpha_{i}}{2\kappa}\mathcal{R}^{(i)} \right).
 \end{equation}

We will name this case as non GR contribution (non-GR-cont).It can be shown that setting $\alpha_1=-1$ is equivalent to fixing $\beta=1$ in Eq. (\ref{eq: F_H_beta}). In Appendix \ref{sec:appendix} we show that in this case  $F(H)$ reads:

\begin{equation}\label{eq:F_H_no_beta}
   F_\delta(H)= H^2+\lambda L^{2(p-1)}H^{2p}e^{\lambda(LH)^{2q}}-H^2 e^{-(\tilde{L}H)^{2s}},
\end{equation}

\noindent where $p$, $q$, and $s$ should meet the following requirements:

\begin{equation}
    p\geq 3, q\geq 1, \text{ and } s\geq 2 \, .
    \label{desig_delta}
\end{equation}

The expression of $F'(H)$ for the $\beta=1$ case is  
\begin{align} 
    F_{\delta}'(H) &= 2H \{ 1 + \lambda  e^{\lambda(LH)^{2q}} (LH)^{2(p-1)} [p  + q \lambda  (LH)^{2q}] \\[3pt]
                  &+ e^{-(\tilde{L}H)^{2s}} [-1 + s (\tilde{L}H)^{2s}] \} \nonumber \,. 
\end{align}

\subsection{Late-time Background Evolution}

From Eq. (\ref{eq: friedmann_eqs_gila}) we can write a differential equation for the Hubble Parameter $H(z)$ as \footnote{To simplify the notation we will not write the subindex 'G/$\beta$/$P$' on the expression of $F(H)$.}
\begin{equation} \label{eq: friedmann_eq_int}
    \frac{dH(z)}{dz} = \frac{\kappa (\rho_{tot} + P_{tot})}{(1+z) F'(H)} \, ,
\end{equation}

where $\rho_{tot} = \rho_{r} + \rho_{m}$ and $F'(H)$ depends on the model that it is considered. The subindex $r$ is for radiation (which includes neutrinos) and $m$ the matter component. Here we recall that in our models the late-time acceleration is provided by the geometry of the model. Besides, from the first Friedmann Eq. (\ref{eq: friedmann_eqs_gila}) a closure relation can be obtained

\begin{equation} \label{eq: clusure_gila}
    \Omega^{GC}_r(z) + \Omega^{GC}_m(z) = 1 \, , 
\end{equation}

where

\begin{equation}
    \Omega^{\rm  GC}_{i,0} = \frac{\rho_{i,0}}{\rho^{\rm GC}_{c,0}} = \frac{\kappa \rho_{i,0}}{3F(H_{0})} \, .
\end{equation}

$\rho^{GC}_c = \frac{3F(H)}{\kappa}$ is the critical density of the GC models. The relation between this critical density and the standard is

\begin{equation}
    \rho^{\Lambda CDM}_c = \frac{F(H)}{H^2}\rho^{GC}_c \, .
\end{equation}

The expression of the Hubble parameter is given by Eq. (\ref{eq: friedmann_eq_int}) and the closure relation (\ref{eq: clusure_gila}).


\section{Observational data} \label{sec:data}
In this section, we describe the cosmological datasets used to constrain GC models considered in this work. 

\subsection{Pantheon Plus + SH0ES (PPS)}

Type Ia supernovae (SNIa) are one of the most luminous events in the Universe, and are considered as standard candles due to the homogeneity of both its spectra and light curves. The distance modulus $\mu$ can be described as,
\begin{equation}
\mu^{th}=25+5 \log_{10}(d_L(z)),
\label{distmod}
\end{equation}

where $d_L$ the luminosity distance

\begin{equation}
d_L(z)=(1+z)\int_0^z\frac{dz'}{H(z')}.
\label{distlum}
\end{equation}
Since the previous expression shows how this last magnitude depends on both the redshift $z$ and the cosmological model [via $H(z)$], it is possible to compare the distance modulus predicted by the theories with the observed ones.

We take into consideration the most recent released Pantheon Plus compilation of type Ia supernovae  \citep{Scolnic:2021amr}. It should be noted that this release includes the option of using low redshift Cepheid data which have been obtained by the SH0ES collaboration. Cepheid data is essential for the calibration of SNIa and consequently for the study of the Hubble tension. With this in mind, we name this data set as Pantheon Plus + SH0ES (PPS). The Pantheon Plus compilation consists of 1701 SNIa with redshift between $0.0012 < z < 2.26$  \footnote{https://github.com/PantheonPlusSH0ES/DataRelease}.  The observed distance modulus estimator can be expressed as
\begin{equation}
\mu=m_B-M+\alpha x_1+\beta c +\Delta_M+\Delta_B,
\label{mu}
\end{equation}
with  $m_B$ being an overall flux normalization, $x_1$ the deviation from the average light-curve shape, and $c$ the mean SnIa B-V color index. Meanwhile, $M$ refers to the absolute B-band magnitude of a fiducial SnIa with $x_1 = 0$ and $c = 0$, $\Delta_B$ refers to a distance correction based on predicted biases from simulations and $\Delta_M$ represent a distance correction based on the mass of the SnIa's  host galaxy. For this SnIa compilation, $\Delta_M$ it is obtained from.
\begin{equation}
\Delta_M=\gamma \times[1+e^{(-(m-m_{\rm step})/\tau)}]^{-1} ,
\end{equation}
where $m_{\rm step}$ and $\tau$ are derived from different host galaxies samples (for details, see~\citet{2018ApJ...859..101S}). $\alpha$, $\beta$, and $\gamma$ are  called nuisance parameters and  determined through a statistical analysis with supernovae data and assuming $\Lambda$CDM. In particular, \citet{2018ApJ...859..101S} obtain for the Pantheon sample  the following values $\alpha=0.0154 \pm 0.006$, $\beta=3.02 \pm 0.06$, and $\gamma=0.053 \pm 0.009$. In addition, other authors, have also estimated the value of the nuisance parameters using the same data set but assuming modified gravity cosmological models \citep{PhysRevD.105.103526, 2020JCAP...07..015N}.  On their results, the estimated nuisance parameters show consistency with those computed by the Pantheon compilation within $1 \sigma$. All those mentioned analyses confirm that the value of the nuisance parameters is independent of the cosmological model. Therefore, in all statistical analyses reported in Sec. \ref{sec:Results} we fix the nuisance parameters to the values published by the Pantheon compilation. Finally, the likelihood of the SNIa data reads
\begin{equation}
\ln {\cal L}_{\rm SN} = - \dfrac{1}{2}\chi^2_{\rm{SN}} = - \dfrac{1}{2} \left( \Delta\mu^{T} \cdot C^{-1} \cdot \Delta\mu \right) \, ,
\end{equation}
where $\Delta\mu=\mu^{th}-\mu^{obs}$ is the difference between the theoretical and the observational distance modulus for each measurement in the compilation, and $C$ is the covariance matrix reported in Ref. \citep{Scolnic:2021amr}.

\subsection{Cosmic chronometers}

Using the differential age evolution of old elliptical passive-evolving objects, one may calculate the Hubble parameter $H(z)$ using the CC approach, which was firstly introduced by \citet{simon05}. Galaxies that formed simultaneously but separated by a small redshift gap are said to be passively evolving, meaning that neither star creation nor contact with neighboring galaxies is occurring. The following expression is used to calculate the Hubble factor $H(z)$ with this method:

\begin{equation}
H(z)=\frac{-1}{1+z}\frac{dz}{dt} \, ,
\end{equation}

where $dz/dt$ can be computed using the ratio $\Delta z/\Delta t$, where $\Delta$ denotes the distinction in the attributes of the two galaxies mentioned above.

In this work we use the most precise available estimates of $H(z)$, which can be found in Refs. \citep{simon05,stern10,moresco12,zhang14,Moresco:2015cya,CC2}. Furthermore, the following likelihood is assumed:
\begin{equation}
\ln {\cal L}_{\rm CC}= - \dfrac{1}{2}\chi^2_{CC} = - \dfrac{1}{2} \sum_i \left(\frac{H\left(z_{i}\right)^{\rm th}-H\left(z_{i}\right)^{\rm obs}}{\sigma_{i}}\right)^{2} ,
\end{equation}
where $H^{\rm th}$ and $H^{\rm obs}$ correspond to the Hubble parameter theoretical prediction and the observable value at redshift $z_i$, respectively. Besides, $\sigma_{i}$ refers to  the standard deviation associated with the dataset $H\left(z_{i}\right)^{\rm obs}$.

\subsection{Globular clusters}
\label{globular}
Globular clusters (GC), dense and nearly spherical groups of stars bound by gravity, are known to host the oldest stars in the Universe. Therefore, the estimation of its ages can be used to constrain the age of the universe (AoU).

The main sequence turnoff (TO) point, is related to the point in the color-magnitude diagram (CMD) at which hydrogen is exhausted in the core, and therefore serves as a primary "clock" for estimating the age of these star clusters. However, the TO point is also sensitive to other parameters such as mass, distance and metallicity among others. Various methods utilize either the TO luminosity or color as key indicators of age, commonly categorized as "vertical" or "horizontal" methods. Vertical methods focus on specific evolutionary sequences observed in a CMD. In contrast, horizontal methods rely on the color of the main-sequence turnoff, rather than its absolute magnitude, to gauge age. In addition, isochrone fitting techniques can be seen as a hybrid approach, as they simultaneously use both magnitude and color to infer age based on the morphology of the fit \citep{2020JCAP...12..002V}. On the other hand, another method to estimate  ages of GC is nucleocosmochronology \citep{2016AN....337..931C}, which does not rely on the TO point. However, all these techniques, although independent of the cosmological model, are subject to various sources of error, which prevents them from being useful to constrain cosmological parameters as the other data described in this section \citep{2018IAUS..334...11C}.

However, we can use this data to estimate a lower limit of the universe's age. We consider an age of 12.2 billion years (Gyr) for the oldest globular cluster, a value that is consistent with the majority recent estimates. Based on this age, we propose a lower limit on the age of the universe at 12.7 Gyr, which includes an additional 0.5 Gyr to account for onset of star formation. We call this last estimate $\rm{AoU^{th}=12.7}$ Gyr. We will use this lower bound in Section \ref{sec:Results} to further constrain the parameter space of the theoretical models considered in this paper. Specifically, we impose a uniform prior on the AoU with a hard lower limit rather than a Gaussian prior, as the latter would unphysically assign non-zero probability to ages below this threshold. We note that the authors of Ref.  \cite{2024PDU....4301401J} also considered the ages of globular clusters and imposed a uniform prior in the same way as in this work. The difference is that they adopted a larger value for the lower bound on the age of the Universe ($\rm{AoU >14.16 \,\,Gyrs}$), making our approach less restrictive.

The value chosen here to limit the age of the universe may seem arbitrary. Therefore, in Section \ref{sec:Results}, we will also discuss how the parameter space changes with reasonable changes in this value.


\section{Methodology} \label{sec:methodology}
In this section, we detail the methodology used to consistently explore the parameter space of the three GC models presented above. Due to  the stiffness of the solutions in certain regions of the parameter space and the non trivial geometry of the prior region given by the globular cluster constraint (see discussion below), a standard MCMC approach is not feasible. We therefore propose a novel and systematic method to test the viability of these models in an organized manner.

Given that the majority of the data sets considered in this work probe the late-time evolution of the universe \footnote{Only the globular cluster constraint provides some information of the early universe. However, this information is not enough to constrain the parameter space sensitive to the the background evolution in the early universe.}, early-time dynamics cannot be constrained. It is therefore not meaningful to include alternative early-universe behavior in the theoretical predictions to be tested. Accordingly, in our cosmological models we fix $\lambda=0$ in Eqs. (\ref{eq: F_H_gila}) and (\ref{eq: F_H_beta}), so that  the remaining parameters characterize possible deviations from $\Lambda$CDM at late times.

For the GC models, the free parameters in the analysis are the absolute magnitude of Type Ia supernovae ($M_\text{abs}$), the Hubble constant ($H_0$), and the parameters already defined $\bar{L}$ and $\beta$.  It is worth noting that Eq. (\ref{eq: clusure_gila}) implies that the reduced matter density $\omega_m= \frac{\kappa}{3 \cdot (100)^2}\rho_{m,0} $  is determined by the values of $\beta$ and $H_0$ as follows:

\begin{equation} \label{eq: GC_omega_m}
    \omega_m = \frac{F(H_0)}{100^2} - \omega_r \, .
\end{equation}

 In all cases, the radiation density is fixed to that observed by COBE $\rho_{r,0} = \frac{3 \cdot (100)^2}{\kappa} 2.47 \times 10^{-5}$ \citep{2009ApJ...707..916F}.

On the other hand, the power-series expansion of $F(H)$ artificially introduces a strong correlation  between $\tilde L$ and $\beta$  (see  Eqs. (\ref{FlateGILAseries}), (\ref{betaGILAcoefficient1}) and (\ref{betaGILAcoefficient2})).
 Therefore, to avoid  non-physical degeneracies,  for each family of models we choose to vary either $\beta$ or $\tilde{L}$, as follows:

 \begin{itemize}
    \item For the GILA model, we fix the energy scale $\tilde{L}=0.90$ in units of $H_0$  and explore the parameter subspace $(M_{abs}, H_0, \beta)$\footnote{We recall that $M_{abs}$ is the absolute B-band magnitude of type Ia supernovae}. This particular value of $\tilde{L}$ is motivated by the requirement that the effects of the theory manifest at late times; therefore, the associated energy scale is fixed close to its present value ($\tilde{L}\sim1$). This choice is based on the analyses  previously carried out for the GILA model in Refs. \cite{Jaime:AUnified} and \cite{Jaime:viability}.
    \item For the GR deformation model, we fix $\beta=10^{-5}$, which is in agreement with the constraint given by \citet{Will:2018bme} from solar system tests.  The free parameter subspace is $(M_{abs},\bar{L},H_0)$. 
    \item For the Geometric Cosmology with no GR contribution, we fix $\beta=1$, which is a subspace that involves the case when the linear Ricci curvature scalar is not considered in the action. The free parameter subspace is $(M_{abs},\bar{L},H_0)$.
 \end{itemize}
For the standard cosmological model $\Lambda$CDM, the free parameters of the model are $(M_\text{abs}, H_0, \omega_m)$. To perform all analyzes, we adopted the priors listed in Table \ref{tab:priors}.

\begin{table}[H]
\caption{\textit{Flat prior ranges on the cosmological parameters of the GC and $\Lambda$CDM models used in all analyses of this work. For the GC models, $\omega_m$ is given by Eq.~(\ref{eq: GC_omega_m}).}}
\label{tab:priors}
\begin{tabularx}{\textwidth}{CCCCC}
\toprule
\textbf{Parameter} & \textbf{GILA} & \textbf{GR-deformation} & \textbf{no GR contr.} & \textbf{$\Lambda$CDM} \\
\midrule
$M_{abs}$ & $[-21,-18.5]$ & $[-21,-18.5]$ & $[-21,-18.5]$ & $[-21,-18.5]$ \\
$H_{0}$ & $[60,80]$ & $[60,80]$ & $[60,80]$ & $[60,80]$ \\
$\beta$ & $[0,12]$ & $\times$ & $\times$ & $\times$ \\
$\log(\bar{L})$ & $\times$ & $[-1,1]$ & $[-1,1]$ & $\times$ \\
$\omega_m$ & $\times$ & $\times$ & $\times$ & $[0.1,0.3]$ \\
\bottomrule
\end{tabularx}
\end{table}

We will fix the values of $(r,s)$ in the GILA model and $s$ in the others to test their viability.  For each such choice,  the corresponding family of cosmological models is parametrized by $H_0,\beta,M_{\rm abs}$. \footnote{$M_{\rm abs}$, the absolute magnitude of SNIa, is not a cosmological parameter. Therefore, for a given choice of 
$(r,s)$ in GILA or $s$ in the remaining models, the cosmological model is determined solely by $H_0,\beta$. Nevertheless, comparison with SNIa data requires treating $M_{\rm abs}$
as a free parameter, so that each model is effectively characterized by $H_0,\beta, M$.}.

\begin{figure}[H]
    \isPreprints{\centering}{} 
    \includegraphics[width=0.7\textwidth]{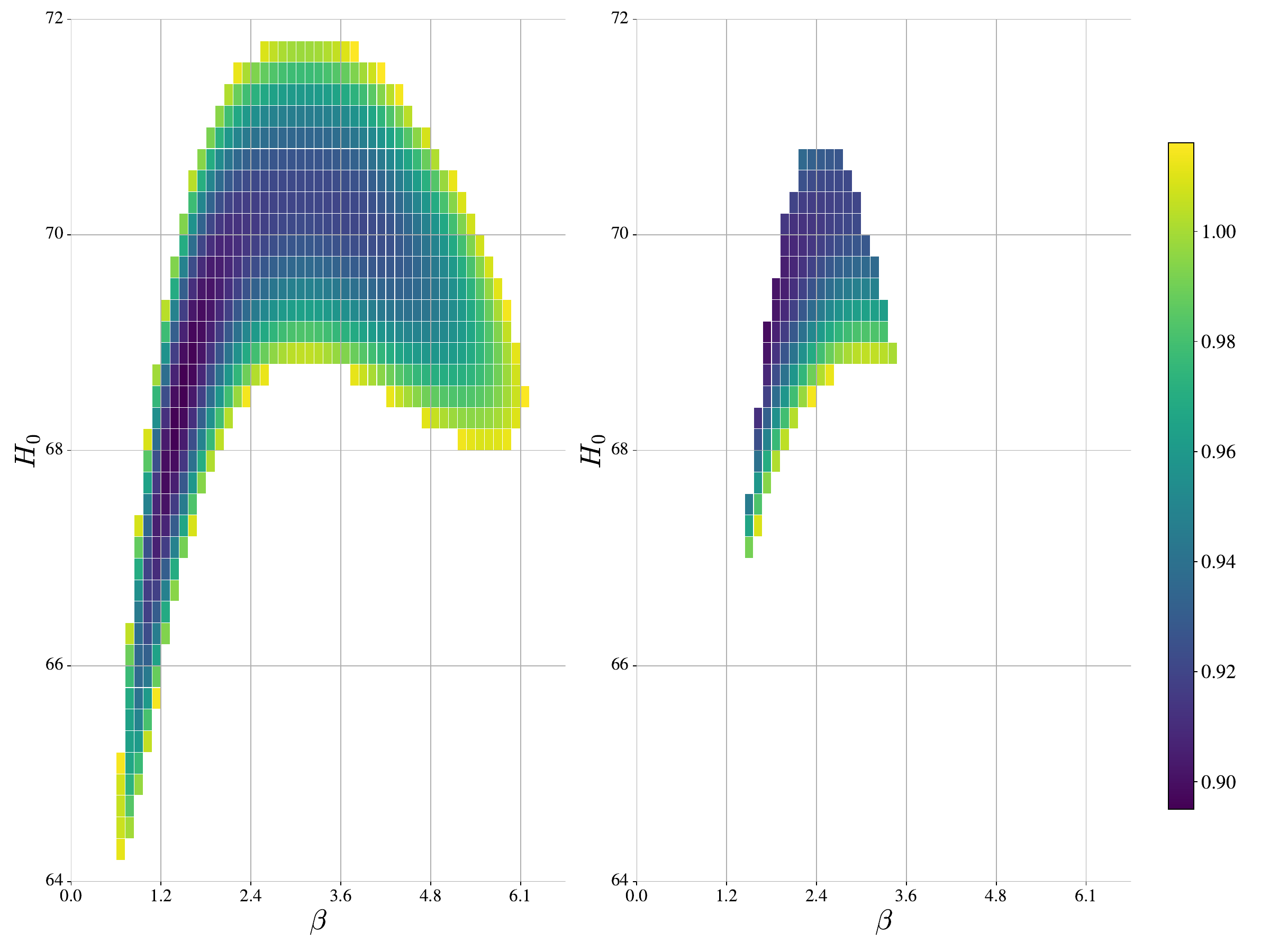}  
    \caption{Left: Heat map for GILA model taking the exponents $(r,s)=(3,4)$ for the energy scale $\tilde{L}=0.90$, using Cosmic Chronometer and type IA supenovae datasets. Right: Same as left but models with $\rm{AoU}<\rm{AoU}^{\rm th}$ are discarded from the grid.}
    \label{fig: heatmap_gila}
\end{figure}

\begin{figure}[H]
    \isPreprints{\centering}{}
    \includegraphics[width=0.7\textwidth]{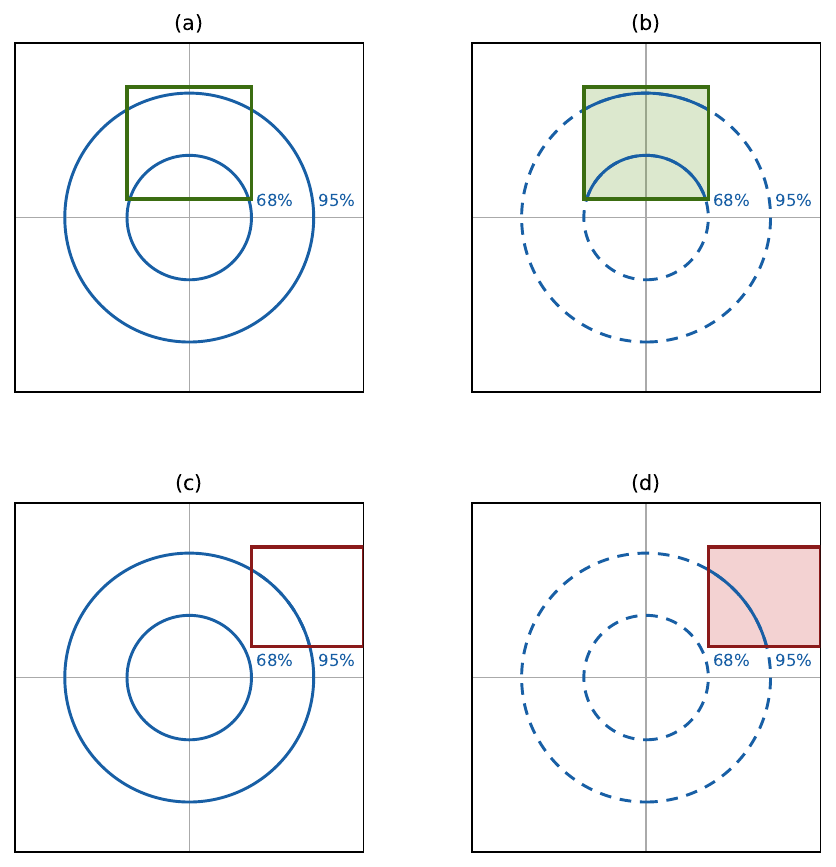} 
    \caption{Illustration of the acceptance region $\mathcal{A}$ in
    parameter space (step 8 in Sect.\ref{sec:methodology}). Each panel shows the 68\% and 95\% credible
    contours of a two-dimensional Gaussian likelihood. The solid green line in panel~(a) and red line in panel~(c) represents $\mathcal{A}$, i.e.\ the
    set of parameter points satisfying the selection condition. 
    Panels~(b) and~(d) show the same configuration but now the non-accepted region is excluded and the shaded region represents $\mathcal{A}$. Panels~(a)--(b) correspond to the
    case where $\mathcal{A}$ overlaps with the 68\% credible region
    (accepted model); panels~(c)--(d) illustrate the case where $\mathcal{A}$ does not (rejected model).}
    \label{fig:filter_gaussian_contours}
\end{figure}

As discussed in Section \ref{globular}, the ages of the oldest globular clusters set a robust lower limit on the age of the Universe (AoU), which is derived quantity of the model's parameters. Accordingly, we impose a uniform prior as follows, 
\begin{equation}
    \pi(\mathrm{AoU}) = 
    \begin{cases}
        0 & \text{if } \mathrm{AoU} < 12.7 \; \text{Gyrs} \\
        1 & \text{otherwise}
    \end{cases}
\end{equation}
A similar approach has been adopted in \cite{2024PDU....4301401J}.
In principle, the posterior distribution could then be computed as the product of the likelihood and this prior \footnote{We note that the uniform prior $[\pi(\mathbf{AoU}) = \Theta(\mathbf{AoU - AoU^{th})}]$ is formally improper, since$[\int_{\mathbf{AoU^{th}}}^{\infty} \pi(\mathbf{AoU}) , d\mathbf{AoU} = \infty.]$. However, this does not affect the validity of our analysis: the essential requirement in Bayesian inference is that the posterior distribution,$[P(\boldsymbol{\theta} \mid D) \propto \mathcal{L}(D \mid \boldsymbol{\theta}) , \pi(\boldsymbol{\theta}),
]$ be proper, i.e.,$[\int P(\boldsymbol{\theta} \mid D) , d\boldsymbol{\theta} < \infty,]$
not necessarily the prior itself. In our case, the likelihood $L(D|\theta)$ is sufficiently constraining, ensuring that the resulting posterior is integrable and well-defined. The use of improper priors is standard practice in Bayesian statistics  and is routinely employed in cosmological parameter estimation.}. However, when we attempted to perform an MCMC analysis using the posterior constructed in this manner, we encountered two main issues: (i) the chains failed to converge due to the complex geometry of the allowed prior region, and (ii) certain regions of parameter space exhibited significant stiffness. These challenges motivated the development of the alternative methodology presented in this work, which is based in MonteCarlo sampling \citep{McElreath2016} (not MCMC) and is, in essence, fully equivalent to an MCMC analysis with an informative prior. 
 \\
The methodology can be divided into three main stages, each comprising several individual steps. In the first stage, we verify that the confidence regions obtained only from the CC and SNIa data sets are mutually consistent for each model under consideration. The second stage addresses the inclusion of the globular cluster constraint on the Age of the Universe  as a prior on the posterior distribution, while the third deals with the sampling of the posterior distribution through Monte Carlo sampling (not MCMC). 
In the first stage, the sampling of the posterior distribution is performed through MonteCarlo sampling (not MCMC). The reason for this is the stiffness in some regions of the parameter space that prevented the MCMC analysis to converge. The first stage comprises the following steps: 
\begin{enumerate}
        \item  For each family of models  (fixed values of $(r, s)$ in the GILA model or $s$ in the remaining models) we construct a 20x20x20 grid over the parameter space and assign to each point of the grid the probability the $e^{-\chi^2_{CC}/2}$  for Monte Carlo sampling.
        \item We sample from this distribution \footnote{As a consistency check, we verified that our results are robust against the choice of grid resolution by rerunning the entire analysis on grids with five times the initial number of points, finding no significant differences.} and then proceed in the usual way to obtain two-dimensional contour plots and one-dimensional marginalized posteriors for each data set separately.
        \item We repeat steps 1 and 2 but using SNIa data instead of CC.
        \item We then verify that the resulting confidence regions from both data sets overlap in some region, ensuring that each model under study can simultaneously account for both data sets with the same set of free parameters. 
\end{enumerate}
Some families of the non-GR-cont case are ruled out at the end of this stage. The second stage consists of the following steps:

\begin{enumerate}[resume]
    \item For each family of models (fixed values of $(r,s)$ in the GILA model or $s$ in the remaining models), we build a low-resolution grid of $20\times20\times20$ in the respective parameter space. For each point on the grid , we compute the total $\chi^2 =\chi^2_{\rm CC}+\chi^2_{\rm PPS}$ (CC and SNIa data).
    \item For each point on the grid, we compute the corresponding Age of the Universe (AoU).
    \item Those points on the grid that meet the requirement $\rm{AoU}<\rm{AoU}^{\rm th}$ are excluded from the grid. Fig. \ref{fig: heatmap_gila} illustrates this step for the $(r,s)=(3,4)$ case.
    \item For each family of models, we evaluate whether there are enough surviving points. In other words, we want to  determine whether
  the cosmological models excluded by the age-of-the-universe condition
fall within the 68\% confidence region of the three dimensional probability distribution defined by the SNIa and CC data. Given $\chi^2_{{(min)}}$ the minimum value of the reduced chi-squared of all surviving points and $\chi^2_{68.3\%}(k)$ the quantile function for $68.3\%$\footnote{The value $\chi^2_{68.3\%}(k)/k$ represents the normalized chi-squared threshold corresponding to the 68.3\% confidence level for $k$ degrees of freedom, where $\chi^2_{68.3\%}(k)$ denotes the inverse cumulative distribution function (or percent point function) of the chi-squared distribution.}, the condition is $\chi^2_{{(min)}} \leq \frac{\chi^2_{68.3\%}(k)}{k}$ \footnote{It is important to note that the 68\% confidence region in one dimension - which is the level conventionally reported for confidence intervals - corresponds to a 39\% confidence region in two dimensions and to 19.9\% in three dimensions. In our analysis, we adopt the 68\% confidence region in the full three-dimensional parameter space, which corresponds to 98.3\% confidence in one dimension. This means that a family of models is discarded only if the minimum $\chi^2$ value among the points that survive the globular cluster age constraint falls outside the 98.3\% confidence level of the one-dimensional marginalized distribution, or equivalently, outside the 82.9\%  confidence level of the two-dimensional marginalized distribution. The adopted criterion is therefore conservative when ruling out a family of models.}. If this holds, we proceed to the following step. If not, the family of models is ruled out on the basis of the globular cluster age constraint. Figs.   \ref{fig: heatmap_gila} and \ref{fig:filter_gaussian_contours} illustrate this step in two dimensions, whereas the procedure itself refers to the full three-dimensional posterior distribution. Tables \ref{tab:gila_exponents_rule_out} and \ref{tab:beta_1_exponents_rule_out} show the results of this procedure when applied to different families that belong to the GILA and GR-deformation models. Note that all GR-deformation models are ruled at this stage because they cannot cannot meet the globular cluster constraint.
\end{enumerate}
In this way, we construct a posterior probability distribution of the GC model parameters $(H_0,\beta, M_{\rm abs})$, which includes as a prior the constraint on the Age of the Universe given by the globular cluster data. 
Note that in step 7, those GC models where the oldest globular clusters did not have enough time to form are discarded from the grid (Fig. \ref{fig: heatmap_gila}), which results in that the sampling of the posterior in next stage will proceed on a reduced region of the parameter space. On the other hand,  two distinct cases can be identified: one in which for a given family of models, the surviving region is located near $\chi^2_{\rm min}$ (the minimum $\chi^2$ value for each family before discarding points from the grid), and another in which it is not.  To quantify this proximity, we employ the criterion outlined  in step 8 and schematized in Fig. \ref{fig:filter_gaussian_contours}, which allows us to rule out a particular family of models (fixed values of $(r,s)$ in the GILA model or $s$ in the remaining models). We recall that the criterion is applied to the 3D posterior distribution while Figs. \ref{fig: heatmap_gila} and \ref{fig:filter_gaussian_contours} are showing an example of a 2D posterior.

The third stage in our analysis is the sampling of the posterior distribution  with Monte Carlo sampling which proceeds through the following steps. We stress that only those families of models  that pass the globular cluster constraints are considered for this stage:

\begin{enumerate}[resume]
    \item We construct the probability distribution for the Monte Carlo sampling, using a narrower grid of $100\times100\times100$ in the regions of the parameter space that were not discarded in the previous stage. For each point on the grid we assign the value of the corresponding likelihood $e^{-\chi^2/2}$, where $\chi^2= \chi^2_{CC}+\chi^2_{SN}$.
    \item We sample from the distribution constructed in the previous step to build the 2D contours plots and 1D posterior distributions.
\end{enumerate}
Since all the sampled points were obtained from the grid points that 
survived the previous stage, this distribution contains information about the Age of  the Globular Clusters in the form of a constraint on the parameter space. On the other hand, we verified that the grid resolution is sufficient to accurately characterize the behavior of the posterior distribution by checking that the reported results do not change if we consider a grid of double resolution.

 We developed a custom Python-based code named \textit{GC-MC} (see \citet{Leizerovich_GC_MC}) , which is built upon the \textit{fR-MCMC} repository \citet{PhysRevD.105.103526}. \textit{GC-MC} performs the numerical integration of Eq. (\ref{eq: friedmann_eq_int}) and estimates the posterior distributions of the parameters using Monte Carlo sampling.

\unskip

\unskip
 


\section{Results} \label{sec:Results}
In this section, we present the results of our analysis.

\subsection{GILA model}
Table \ref{tab:gila_exponents_rule_out} shows the coefficients $r$ and $s$ that characterize the ten particular cases of the GILA model considered in this work. Additionally, the mentioned table indicates which models are ruled out because their predicted Age of the Universe is below the estimated value from Globular Clusters (see Section \ref{sec:data}). As explained in Section \ref{sec:methodology}, in the second stage we build a grid in the parameter space for fixed values of the energy scale $\tilde{L}$ and exponents $r$ and $s$.  Next, we discuss an example case to illustrate steps 5-7 of the methodology. Fig. \ref{fig: heatmap_gila} shows a heatmap (taking the exponents $(r,s)=(3,4)$) for the probability distribution for the parameters $(\beta,H_0)$, fixing $M_{abs}$ on its best fit value, together with the same plot, but excluding the points for which the Age of the Universe is below our estimated value in Section \ref{sec:data}. Although this Figure does not exactly reflect the methodology since we are fixing the value of $M_{abs}$, we believe that it is useful to illustrate the described steps 5-7.  

\begin{table}[H]
\caption{\textit{Viability of the GILA model for different sets of exponents (r,s). Models with $\chi^2_{k_{(\min)}} \leq \frac{\chi^2_{68.3\%}(k)}{k}$ are marked with a checkmark ($\checkmark$), while those that do not satisfy this condition are marked with a cross ($\times$). For the former, we also show the reduced $\chi^2$ and the results of the AIC/BIC comparison model test, using $\Lambda$CDM as the baseline model. For comparison we also provide $\chi^2_\nu (\Lambda {\rm CDM}): 0.877 $}
\label{tab:gila_exponents_rule_out}}
\begin{tabularx}{\textwidth}{CCCC}
\toprule
\textbf{$(r,s)$} & \textbf{GILA} & $\chi^2_\nu$ & $\Delta$ AIC / $\Delta$ BIC\\
\midrule
$(3,1)$ & $\times$ & $-$  & $-$ \\
$(3,2)$ & $\times$ & $-$ & $-$ \\
$(3,4)$ & $\checkmark$ & $0.899$ & $36.9$\\
$(3,5)$ & $\checkmark$ & $0.899$ & $37.9$\\
$(3,6)$ & $\checkmark$ & $0.9$ & $39.1$\\
$(4,1)$ & $\times$ & $-$ & $-$\\
$(5,1)$ & $\times$ & $-$ & $-$\\
$(6,1)$ & $\times$ & $-$ & $-$\\
$(7,1)$ & $\times$ & $-$ & $-$\\
$(8,1)$ & $\times$ & $-$ & $-$\\
\bottomrule
\end{tabularx}
\end{table}

Table \ref{tab:results_GILA} shows the constraints on the parameters for the viable GILA models that are obtained from the sampling of the posterior probability which in turn is constructed using a denser grid (see step 9 described in Section \ref{sec:methodology}). 

\begin{figure}[H]
    \isPreprints{\centering}{} 
    \includegraphics[width=0.7\textwidth]{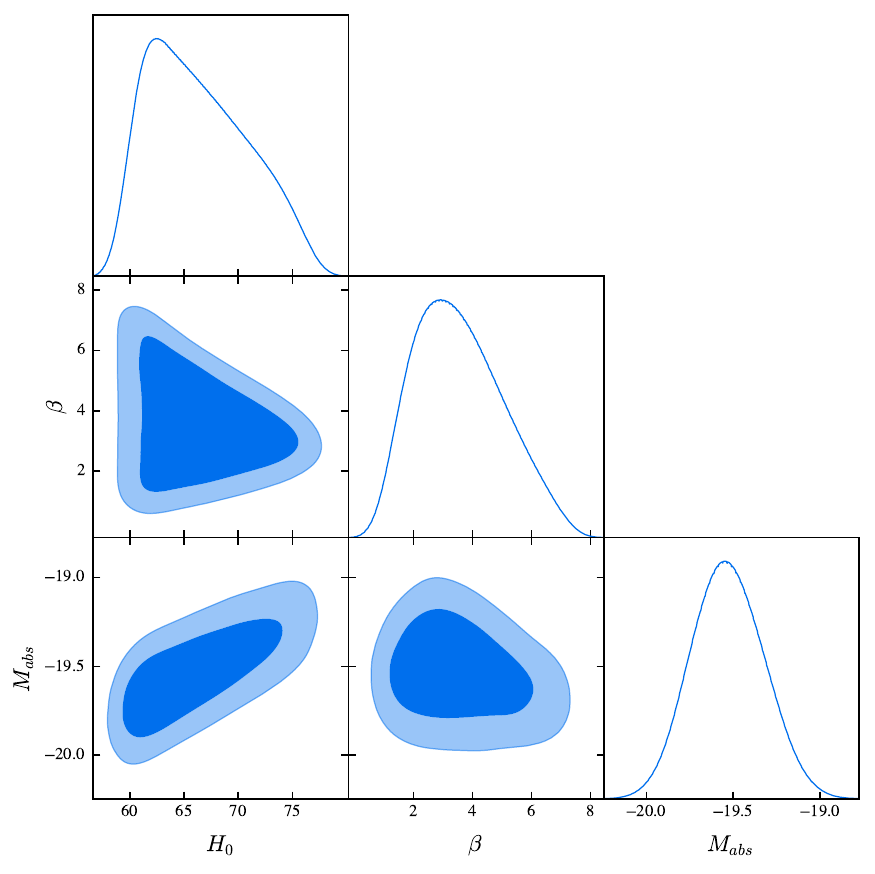}
    \caption{Results of the statistical analysis with Cosmic Chronometer and type IA supenovae for the GILA model, taking the exponents $(r,s)=(3,5)$. The energy scale is fixed $\tilde{L}=0.90$. The darker and brighter regions correspond to the $68\%$ and $95\%$ confidence regions respectively. The plots on the diagonal show the posterior probability density for each of the free parameters of the model.}
    \label{fig: getdist_gila}
\end{figure}

Fig. \ref{fig: getdist_gila} shows the 1D posterior probabilities together with the 2-dimensional confidence contours for the particular case $(r,s)=(3,5)$. For this model, both values related to the $H_0$ tension are allowed. Looking at the posterior contour for $H_0$, it can be seen that the data are not able to constrain $H_0$ beyond what is imposed by the prior. However, the data are informative enough to constrain higher values of $H_0$, effectively setting limits on the parameter space. On the other hand, the parameter $\beta$ does not show correlations with $H_0$ or $M_{abs}$. Besides, the posterior contour is informative with respect to the prior on this parameter, as shown in Table \ref{tab:results_GILA}. We stress that one of the main achievements of this work lies in constraining the values of parameter $\beta$ with observational data. On the other hand, we note that the results for the other viable GILA models are similar to the example case as shown in Table \ref{tab:results_GILA}. 

Furthermore, we test different threshold values $\rm AoU^{th}$ in order to test the consistency of our analysis. In particular, we see that taking a threshold $\rm AoU^{th} $ between $\rm 12\,Gyr$ and $\rm 13\,Gyr$ does not change our results.

Finally, Table~\ref{tab:gila_exponents_rule_out} presents the results of applying model comparison criteria using $\Lambda$CDM as the baseline model. We also report the minimum reduced $\chi^2$ values for both the surviving GILA models and $\Lambda$CDM, which confirm that the fits provide an adequate description of the data. The results from model comparison criteria (also shown in Table ~\ref{tab:gila_exponents_rule_out}) indicate that $\Lambda$CDM is statistically favored by the data over the GILA models that survived the globular cluster constraint.

\unskip

\subsection{GR-deformation}
We considered six particular models of the GR-deformation family, which are characterized by the value of the coefficient s. Table \ref{tab:beta_1_exponents_rule_out} shows that none of the considered cases predict an Age of the Universe consistent with the estimated one from Globular Clusters and therefore these models can be ruled out.

\subsection{Geometric Cosmology with non GR contribution}

We considered five specific cases within this family, employing different exponents $s$. Models with $s = \{2, 3, 4\}$  are ruled out in step 4 (see Section \ref{sec:methodology}), as the confidence intervals for the free parameters obtained from the CC dataset are inconsistent with those derived from the PPS analysis. In contrast, the cases with $s=\{5,6,7\}$ are excluded in step 8, as these models fail to satisfy the age of the Universe constraint established in the previous section.


\begin{table}[H]
\caption{Evaluations of the viability of the GR-deformation model for different exponents $s$. The cross indicates that the evaluated GR-deformation models do not pass the test $\chi^2_{k_{(min)}} \leq \frac{\chi^2_{68.3\%}(k)}{k}$.}
\label{tab:beta_1_exponents_rule_out}
\begin{tabularx}{\textwidth}{CCCC}
\toprule
\textbf{$s$} & \textbf{GR-Deformation} \\
\midrule
$2$ & $\times$ \\
$3$ & $\times$ \\
$4$ & $\times$ \\
$5$ & $\times$ \\
$6$ & $\times$ \\
$7$ & $\times$ \\
\bottomrule
\end{tabularx}
\end{table}

\begin{table}[H]
\caption{Results from the statistical analysis using data from CCs and SnIa. For each parameter, we present the mean value and the 68\% (95\%) confidence levels.}
\label{tab:results_GILA}
\begin{tabularx}{\textwidth}{CCCCC}
\toprule
\textbf{$\Lambda$CDM} &  & \textbf{$M_{abs}$} & \textbf{$\omega_m$} & \textbf{$H_0$} \\
\midrule
 &  & $-19.289_{-0.025(0.050)}^{+0.024(0.047)}$ 
 & $0.1636_{-0.0095(0.019)}^{+0.0099(0.020)}$ 
 & $72.20_{-0.8(1.7)}^{+0.9(1.7)}$ \\
\midrule
\textbf{GILA model} & \textbf{Exponents $(r,s)$} & \textbf{$M_{abs}$} & \textbf{$\beta$} & \textbf{$H_0$} \\
\midrule
 & $(3,4)$ & $-19.56_{-0.2(0.35)}^{+0.13(0.3)}$ 
 & $2.92_{-1.7(2.1)}^{+0.6(2)}$ 
 & $64.13_{-4.1(4.1)}^{+1.3(5.2)}$ \\
 & $(3,5)$ & $-19.50_{-0.24(0.37)}^{+0.19(0.44)}$ 
 & $3.722_{-2(2.8)}^{+1.2(3.1)}$ 
 & $67.4_{-7.4(7.4)}^{+2.5(9.4)}$ \\
 & $(3,6)$ & $-19.48_{-0.28(0.41)}^{+0.17(0.43)}$ 
 & $4.991_{-3.3(4)}^{+1.8(4.3)}$ 
 & $68.7_{-8.7(8.7)}^{+3(9.9)}$ \\
\bottomrule
\end{tabularx}
\end{table}


\section{Discussion and Conclusions} \label{sec:conclusions}

In this work, we have explored the late-time evolution of Geometric Cosmology theories, which provide an explanation for the late-time acceleration of the Universe without requiring a cosmological constant or an additional dark component in the energy-momentum tensor. For this study, we focussed on three cases of interest: the GILA model, the GR-deformation scenario, and models with non-GR contribution. We further restricted our analysis by assuming an exponential ansatz for $F(H)$, the characteristic  function that appears in the modified Friedmann equations. 
Also, due to the number of free parameters and the strong degeneracy between them, we restricted our analysis to selected examples within each class of models to test their viability.

An interesting question that arises from our analysis is how critical the globular cluster data are in discriminating among models.  To address this point, we repeated the Monte Carlo sampling for each model considered in this work  using both CC and SNIa data, but without imposing the prior on the age of the Universe, and then applied the model comparison criteria. We found that the results remain unchanged: the $\Lambda$CDM model is still preferred by the data. However, we wish to emphasize that the methodology developed in this work for incorporating the globular cluster constraint can itself be considered one of the main results, as it can be applied to other $F(H)$ choices within the GC models or to other alternative theories of gravity for which the use of this constraint may be more critical than in the case studied here.

We now summarize the conclusions obtained from the statistical analyzes of the models considered here with the observational data.
\begin{enumerate}
 \item Some of the non-GR-contribution
 models analyzed in this paper can be ruled out because the parameter 
intervals obtained from CC are inconsistent with those obtained from SNIa.
        \item The GR-deformation models considered in this paper  and  the non-GR contribution models that were not discarded due to inconsistency between CC and SNIa constraints   can be ruled out because their predicted age of the Universe is lower than the age inferred from globular clusters.
       
        \item For the GILA model, we find three particular choices of the coefficients $r$ and $s$ that can explain the CC and PPS data and also predict an Age of the Universe consistent with globular cluster data. For these models, we were able to estimate confidence intervals for the free parameters. However, model comparison criteria show that $\Lambda$CDM is favored by the data against these models.
\end{enumerate}
This work presents the first systematic test of this class of models with recent observational data, incorporating the constraints derived from the ages of the oldest globular clusters — a test that is frequently overlooked in the literature. The inclusion of these data, combined with the mathematical structure of the modified Friedmann equation, required the development of a dedicated methodology, which represents an additional contribution of this study. Although the models examined here — restricted to a specific functional form of $F(H)$ - are either excluded by the data or statistically disfavored relative to $\Lambda$CDM, the present work establishes a robust framework for testing this class of models, rules out a particular choice of $F(H)$, and opens the path for future studies exploring alternative functional forms.


\authorcontributions{Conceptualization, L.J. and S.L.; methodology, M.L. and S.L.; software, M.L.;  formal analysis, L.J. and G.A.;  data curation, M.L and S.L.; writing---original draft preparation, M.L. and L.J. and S.L and G.A.; writing---review and editing, M.L. and L.J. and S.L and G.A. . All authors have read and agreed to the published version of the manuscript.}

\funding{
The authors  acknowledge the use of the supercluster- MIZTLI of UNAM through project  LANCAD-UNAM-DGTIC-449 and thank the people of DGTIC-UNAM for technical and computational support. 
G.A. and L.G.J. acknowledge financial support from SECIHTI-SNII. 
L.G.J. acknowledges financial support from CONAHCYT (grant 140630). 
S.L. and M.L. are supported by CONICET grant PIP 11220200100729CO and UBACYT grant 20020170100129BA. LGJ gratefully acknowledges the fellowship from the Secretaría de Ciencias, Humanidades, Tecnología e Innovación (SECIHTI, formerly CONAHCYT) fellowship, CVU: 45775.
}

\dataavailability{The data and analysis code supporting the results of this study are publicly available at GitHub: \url{https://github.com/matiasleize/GC_MC}. The repository contains the scripts used for the statistical analyses as well as the datasets employed in this work.}

\acknowledgments{
The authors would like to acknowledge the computational resources provided by the Miztli supercomputer (DGTIC, UNAM), which were essential for performing the numerical analyses presented in this work. The authors are grateful to Hermano Velten for highlighting the importance of the globular-cluster age constraint on the age of the Universe.
}

\conflictsofinterest{The authors declare no conflicts of interest.}

\appendix

\section{Exponential convergence} \label{sec:appendix}

In this Appendix, we show the exponential convergence for the models that we analyze in this paper: GILA and GR-deformation. Let us start with the general expression for the $F(H)$ 
 infinite series expansion of Eq. (\ref{eq: fdeh})

\begin{equation}\label{eq:appfdeh1}
    F(H)=(1+\alpha_1)H^2+\sum_{i=3}^{\infty}\alpha_{(i)}H^{2i}.
\end{equation}

We can split the summation into two by taking $\alpha_{(i)}=\tilde{\alpha}_{(i)}+\hat{\alpha}_{(i)}$, 

\begin{eqnarray}\nonumber
    &&F(H)=(1+\alpha_1)H^2+\sum_{i=3}^{\infty}\tilde{\alpha}_{(i)}H^{2i}+\sum_{i=3}^{\infty}\hat\alpha_{(i)}H^{2i},\\ \label{fsplit}
    &&=(1+\alpha_1)H^2+\sum_{m=3}^{\infty}\tilde{\alpha}_{(m)}H^{2m}+\sum_{n=3}^{\infty}\hat\alpha_{(n)}H^{2n}.
\end{eqnarray}

\noindent where we considered that the summation over $i$ runs independently in both summations such that we changed the index $i$ for $m$ and $n$ without loss of generality.

We will use the first summation for the early-time acceleration mechanism and a tilde over the coefficients for the early-time, and the last summation for the late-time accelerated era with a hat over the coefficients for the late-time, so we define:

\begin{eqnarray}
    F_{\rm Early}(H)&=&\sum_{m=3}^{\infty}\tilde\alpha_{(m)}H^{2m},\\ \label{FlateGILAseries}
    F_{\rm Late}(H)&=&\sum_{n=3}^{\infty}\hat{\alpha}_{(n)}H^{2n},\\
\end{eqnarray}

Let us start with the $F_{\rm Early}(H)$. If we take the first non-zero coefficient $\tilde{\alpha}_{m}$, and name it $\tilde{\alpha}_{(p)}$, i.e. $p={\rm min}\{m|\tilde{\alpha}_{(m)}\neq 0\}$, we can factorize it from the summation:

\begin{eqnarray}
    F_{\rm Early}(H)&=&\tilde{\alpha}_{(p)}H^{2p}\sum_{m=p}^{\infty}\frac{\tilde\alpha_{(m)}}{\tilde{\alpha}_{(p)}}H^{2(m-p)},\\
    &=&\tilde{\alpha}_{(p)}H^{2p}\sum_{j=0}^{\infty}\tilde{\xi}_{(j)}H^{2j}.
\end{eqnarray}

\noindent where we defined $j=m-p$, and $\tilde{\xi}_{(j)}=\tilde{\alpha}_{(m)}/\tilde{\alpha}_{(p)}$.  
Clearly, $p\geq 3$, and $m\geq p$. 

Analogously, for the late-time $F_{\rm Late}(H)$ we get:

\begin{eqnarray}
    F_{\rm Late}(H)&=&\hat{\alpha}_{(r)}H^{2r}\sum_{n=r}^{\infty}\frac{\hat\alpha_{(n)}}{\hat{\alpha}_{(r)}}H^{2(n-r)},\\ \label{factorizeLate}
    &=&\hat{\alpha}_{(r)}H^{2r}\sum_{k=0}^{\infty}\hat\xi_{(k)}H^{2k}.
\end{eqnarray}

At this level, we do not have a prescription to fix the $\alpha_{(i)}$ coefficients, but instead of taking random values for simplicity for $\tilde\xi_{(j)}$ and $\hat\xi_{(k)}$, we can consider only the case where $j$ and $k$ are a multiple of a certain value $q$ and $s$, respectively, i. e.

\begin{eqnarray}\label{Fearlyj0}
    F_{\rm Early}(H)&=&\tilde{\alpha}_{(p)}H^{2p}\sum_{j=0}^{\infty}\tilde\xi_{(j\cdot q)}H^{2j\cdot q},\\ \label{smultiples1}
    F_{\rm Late}(H)&=& \hat{\alpha}_{(r)}H^{2r}\sum_{k=0}^{\infty}\hat\xi_{(k\cdot s)}H^{2k\cdot s}.
\end{eqnarray}

Before continuing, let us first take a slightly different expression for the same $F(H)$ given in Eq. (\ref{eq:appfdeh1}). Instead of factorizing $\hat{\alpha}_{(r)}H^{2r}$ from the late acceleration $F_{\rm Late}(H)$ in Eq. (\ref{factorizeLate}), we factorize $\hat{\alpha}_{(r=1)}H^{2}$, i.e.

\begin{eqnarray}
    F_{\rm Late}(H)&=&\hat{\alpha}_{(r=1)}H^{2}\sum_{n=3}^{\infty}\frac{\hat\alpha_{(n)}}{\hat{\alpha}_{(r=1)}}H^{2(n-1)},\\ \label{factorizeLater1}
    &=&\hat{\alpha}_{(r=1)}H^{2}\sum_{k=2}^{\infty}\hat\xi_{(k)}H^{2k}.
\end{eqnarray}

Notice that the summation in Eq. (\ref{factorizeLater1}) starts at $k=2$ instead of $k=0$ in order to maintain $H^2\times H^{2k}|_{k=2}=H^6$ as the lowest $H^{2n}$ term in the series expansion.

If we want to consider only multiples of a fixed value $s$ as was done in (\ref{smultiples1}), then

\begin{equation}\label{later1}
    F_{\rm Late}(H)= \hat{\alpha}_{(r=1)}H^{2}\sum_{k=1}^{\infty}\hat\xi_{(k\cdot s)}H^{2k\cdot s},
\end{equation}

\noindent where now $s\geq 2$. 

Let us remark that the construction given until now, from Eq. (\ref{fsplit}), is considering the following structure:

\begin{equation}
    F(H)=(1+\alpha_1)H^2+F_{\rm Early}(H)+F_{\rm Late}(H).
\end{equation}

However, if $\alpha_1\neq 0$, we can include the term $\alpha_1 H^2$ into the summation in the $F_{\rm Late}(H)$. In this case, the only change in the expression for the late acceleration era is to start at $k=0$, and replacing  $\hat{\alpha}_{(r=1)}$ for $\alpha_1$, but maintaining the condition $s\geq 2$ to avoid the emergence of an $H^4$ (that belongs to the Gauss-Bonnet Lagrangian density), i.e.

\begin{equation}\label{laterbeta}
    F_{{\rm Late},\beta}(H)= \hat{\alpha}_{(r=1)}H^{2}\sum_{k=0}^{\infty}\hat\xi_{(k\cdot s)}H^{2k\cdot s}, \, (s\geq 2).
\end{equation}

and

\begin{equation}\label{fbetaconstruction}
    F_\beta(H)=H^2+F_{\rm Early}(H)+F_{{\rm Late},\beta}(H),
\end{equation}

\noindent where we added a label $\beta$ in Eq. (\ref{fbetaconstruction}) because this is the case for the GR-deformation $F_\beta(H)$ in section \ref{GRdeformation}.

An exploratory analysis performed in a previous work \citet{Jaime:AUnified} indicates that exponential solutions may explain the current type Ia supernovae datasets. Therefore, in order to obtain exponential solutions for $F_{\rm Early}(H)$ the coefficients $\tilde{\alpha}_{(p)}$ and $\hat{\xi}_{(j\cdot k)}$ must obey

\begin{eqnarray}\label{betaGILAcoefficient1}
    \tilde{\alpha}_{(p)}&=&\lambda L^{2(p-1)},\\ \label{betaGILAcoefficient2}
    \hat{\xi}_{(j\cdot q)}&=&\lambda^jL^{2j\cdot q}/j!,
\end{eqnarray}

\noindent and Eq. (\ref{Fearlyj0}) reads

\begin{eqnarray}\label{fearlyseriesExp}
    F_{\rm Early}(H)&=&\lambda L^{2(p-1)}H^{2p}\sum_{j=0}^{\infty}\frac{\lambda^j}{j!}(LH)^{2j\cdot q},\\ \label{fearlyconvergExp}
   \Rightarrow F_{\rm Early}(H) &=&\lambda L^{2(p-1)}H^{2p} e^{\lambda(LH)^{2q}}.
\end{eqnarray}

\noindent where $p\geq 3$, and $q\geq 1$.

Similarly, for the $F_{\rm Late}(H)$ in Eq. (\ref{smultiples1}), conditions for coefficients $\hat{\alpha}_{(r)}$ and $\hat{\xi}_{(k\cdot s)}$ are

\begin{eqnarray}
    \hat{\alpha}_{(r)}&=&-\beta \tilde{L}^{2(r-1)},\\
    \hat{\xi}_{(k\cdot s)}&=&(-1)^{k}\beta^k\tilde{L}^{2k\cdot s}/k!,
\end{eqnarray}

\noindent then Eq. (\ref{smultiples1}) reads

\begin{eqnarray}\label{flateserieExp}
    F_{\rm Late}(H)&=&-\beta \tilde{L}^{2(r-1)}H^{2r}\\ \nonumber
    &&\times \sum_{k=0}^{\infty}\frac{(-1)^{k}\beta^k}{k!}(\tilde{L}H)^{2k\cdot s},\\\label{flateconvergExp}
  \Rightarrow  F_{\rm Late}(H)  &=&-\beta \tilde{L}^{2(r-1)}H^{2r} e^{-\beta(\tilde{L}H)^{2s}},
\end{eqnarray}

\noindent where $r\geq 3$ and $s\geq 1$.

Now, we are ready to make an explicit construction of the exponential models that we used in this work for the different theories we explored in the main text. 

\subsection{The GILA model} 

The GILA model belongs to the family of theories given in action (\ref{actionGC}), but with two considerations: (1) $\alpha_1=0$, and (2) the infinite series expansion $F_\gamma(H)$ converges to exponential functions. Therefore, the action for GILA models is

\begin{equation}
    S_{\gamma} = \frac{1}{2\kappa}\int d^4x  \sqrt{-g}\left[ R + \sum^{\infty}_{i=3}\alpha_{i}\mathcal{R}^{(i)}\right] \, ,
\end{equation}

\noindent and the $F_\gamma(H)$ is given by Eqs. (\ref{fearlyseriesExp}) and (\ref{flateserieExp})

\begin{multline}
F_\gamma (H)=H^2+    \lambda L^{2(p-1)}H^{2p}\sum_{j=0}^{\infty}\frac{\lambda^j}{j!}(LH)^{2j\cdot q}\\
-\beta \tilde{L}^{2(r-1)}H^{2r}\sum_{k=0}^{\infty}\frac{(-1)^{k}\beta^k}{k!}(\tilde{L}H)^{2k\cdot s}
\end{multline}

which converges to (\ref{fearlyconvergExp}) and (\ref{flateconvergExp})

\begin{multline} 
   F_{\gamma}(H)=H^2+ \lambda L^{2(p-1)}H^{2p} e^{\lambda(LH)^{2q}}
   -\beta \tilde{L}^{2(r-1)}H^{2r} e^{-\beta(\tilde{L}H)^{2s}}.
\end{multline}

\noindent with conditions $p\geq 3$, $q\geq 1$, $r\geq 3$, and $s\geq 1$.

\subsection{GR-deformation}

The action for the GR-deformation theory is given by the action of Eq. (\ref{actionGC}) when $-1<\alpha_1 <0$ 

\begin{equation}
    S_{\beta} = \frac{1}{2\kappa}\int d^4x  \sqrt{-g}\left[(1+\alpha_1) R + \sum^{\infty}_{i=3}\alpha_{i}\mathcal{R}^{(i)}\right] \, .
\end{equation}

In this case, the $F_\beta(H)$ is given by the same equation for the early-time (\ref{fearlyseriesExp}), but Eq. (\ref{laterbeta}) for late-time, with decomposition given in (\ref{fbetaconstruction}), and condition $s\geq 2$.

\begin{multline}
F_\gamma (H)=H^2+    \lambda L^{2(p-1)}H^{2p}\sum_{j=0}^{\infty}\frac{\lambda^j}{j!}(LH)^{2j\cdot q}\\
-\beta H^{2}\sum_{k=0}^{\infty}\frac{(-1)^{k}\beta^k}{k!}(\tilde{L}H)^{2k\cdot s},\,
(s\geq 2).
\end{multline}

\noindent which converges to Eq. (\ref{eq: F_H_beta}):

\begin{equation}
   F_\beta(H)= H^2+\lambda L^{2(p-1)}H^{2p}e^{\lambda(LH)^{2q}}-\beta H^2 e^{-\beta(\tilde{L}H)^{2s}},
\end{equation}

\noindent with conditions $p\geq 3$, $q\geq 1$ and $s\geq 2$.

\subsection{Non-GR contribution}

The case when there is no GR contribution is when $\alpha_1=-1$ in action (\ref{actionGC}):

\begin{equation}
    S_{\delta} = \frac{1}{2\kappa}\int d^4x  \sqrt{-g}\left[ \sum^{\infty}_{i=3}\alpha_{i}\mathcal{R}^{(i)}\right] \, .
\end{equation}

In this case, the $F_\delta(H)$ series expansion is given by the same equation for the early-time (\ref{fearlyseriesExp}), but Eq. (\ref{later1}) for late-time:

\begin{multline}
 F_\gamma (H)=H^2+    \lambda L^{2(p-1)}H^{2p}\sum_{j=0}^{\infty}\frac{\lambda^j}{j!}(LH)^{2j\cdot q}\\    -\beta H^{2}\sum_{k=1}^\infty \frac{(-1)^{k}\beta^{k-1} }{k!}(\tilde{L} H)^{2s\cdot k}, \, (s\geq 2),
\end{multline}

\noindent which converges to

\begin{equation}\label{Fdelta-app}
   F_{\delta}(H)= H^2+\lambda L^{2(p-1)}H^{2p}e^{\lambda(LH)^{2q}}-H^2e^{-\beta(\tilde{L}H)^{2s}}.
\end{equation}

The last Eq. (\ref{Fdelta-app}) is the same as the GR-deformation theory when $\beta=1$. This is the case analyzed in section \ref{secc:noGR}.

\reftitle{References}
\isAPAStyle{}{
\bibliography{references}
}

\end{document}